# Symmetry-breaking of turbulent flow due to asymmetric vortex shedding in periodic porous media

**Vishal Srikanth[1] and Andrey V. Kuznetsov[1,a)]**

**AFFILIATIONS**

[1]Department of Mechanical and Aerospace Engineering, North Carolina State University, Raleigh, NC 27695, USA

[a)]**Author to whom correspondence should be addressed:** avkuznet@ncsu.edu

**ABSTRACT**

In this paper, we report new insight into a symmetry-breaking phenomenon that occurs for turbulent flow in periodic porous media composed of cylindrical solid obstacles with circular cross-section. We have used Large Eddy Simulation to investigate the symmetry-breaking phenomenon by varying the porosity (0.57-0.99) and the pore scale Reynolds number (37-1,000). Asymmetrical flow distribution is observed in the intermediate porosity flow regime for values of porosities between 0.8 and 0.9, which is characterized by the formation of alternating low and high velocity flow channels above and below the solid obstacles. These channels are parallel to the direction of the flow. Correspondingly, the microscale vortices formed behind the solid obstacles exhibit a bias in the shedding direction. The transition from symmetric to asymmetric flow occurs in between the Reynolds numbers of 37 (laminar) and 100 (turbulent). A Hopf bifurcation resulting in unsteady oscillatory laminar flow marks the origin of a secondary flow instability arising from the interaction of the shear layers around the solid obstacle. When turbulence emerges, stochastic phase difference in the vortex wake oscillations caused by the secondary flow instability results in flow symmetry breaking. We note that symmetry breaking does not occur for cylindrical solid obstacles with square cross-section due to the presence of sharp vertices in the solid obstacle surface. At the macroscale level, symmetry-breaking results in residual transverse drag force components acting on the solid obstacle surfaces. Symmetry-breaking promotes attached flow on the solid obstacle surface, which is potentially beneficial for improving transport properties at the solid obstacle surface such as convection heat flux.

## I. INTRODUCTION

Turbulent flow is encountered in numerous scenarios for flow in porous media with Reynolds number values as low as 100[1]. For example, turbulence is observed for the flow through industrial scale heat exchangers[2], cooling systems for electronics[3], and forest fire modeling[4]. The solid obstacles that compose the porous medium in the above flow scenarios are often cylindrical in shape and the space between them is close enough that the vortices that are formed behind the solid obstacles impinge on the neighboring solid obstacle that is located downstream. Turbulent flow is characterized by flow instability and stochasticity, which result in unique flow behavior in porous media due to the interaction of the flow with numerous solid obstacles. In addition to its theoretical value, understanding the characteristics of turbulent flow instabilities in porous media has immense practical utility in developing robust mathematical models for environmental and energy applications, as discussed earlier.

Recently, many studies have focused on investigating turbulence statistics in porous media, fueled by the advances in high performance computing and high-fidelity modeling approaches. These studies have highlighted the fundamental characteristics of turbulence in periodic porous media, such as the turbulence length and time scale limitations imposed by the pore geometry[5–9], and the absence of macroscale turbulent structures[10,11] with the possible exception of extremely porous media (>98%)[12]. At the microscale level, turbulent flow structures are generated as microscale vortices behind the solid obstacles that compose the porous medium[13]. The vortices are advected downstream where the confining geometry of the pores causes its dissipation either in the pore space or upon impingement on a neighboring solid obstacle. Generalizing turbulent flow behavior by considering different solid obstacle geometries, the properties of microscale turbulence and the related macroscale flow variables, such as drag force and heat flux, are determined by the distribution of commonly encountered flow features inside the pores. For example, flow recirculation behind solid obstacles decreases the local shear stress and heat flux, whereas shear flow conditions around the solid obstacle increase viscous drag and local heat flux. While there is a strong dependence of the microscale turbulent flow distribution on the geometry of the solid obstacles (or pore space), these fundamental qualitative observations have been reported for a wide range of geometries and Reynolds numbers[5,7,13,14].

It is interesting to note that the microscale flow behavior in porous media deviates from the expected flow behavior observed in canonical flow configurations when flow instabilities interact with the solid obstacle surfaces in the confining pore space. Flow instabilities caused by the interaction of the flow with numerous solid obstacles have not yet been studied in detail, especially in the turbulent flow regime. A few researchers have reported the occurrence of different types of bifurcations and instabilities of flow in porous media. For laminar flow in a 2D periodic porous medium, the formation of asymmetrical vortices was observed when square solid obstacles were arranged in a staggered manner[15]. This numerical study reported critical values of porosity (0.561-0.704) and Reynolds number (250-500) for a pitchfork bifurcation in the behavior of the vortex flow patterns with 3 possible solutions – symmetric and asymmetric (2 modes) vortex rotation. Another numerical study of laminar flow in porous media discovered the occurrence of a Hopf bifurcation for both square and circular solid obstacles leading to unsteadiness of the flow, harmonic oscillations of the drag force, and increased tortuosity of the flow streamlines[16,17]. The Hopf bifurcation is observed for a wide range of porosities from 0.15 to 0.96 with the critical Reynolds number for bifurcation at a porosity of 0.82 equal to 49. This study is relevant to our present work as we show in section III that the Hopf bifurcation for laminar flow is a causative factor that precedes the symmetry-breaking for turbulent flow. We note that the unsteadiness of the flow at laminar flow Reynolds numbers is also reported in Forslund et al.[18]. We would also like to note that we have observed flow symmetry-breaking in the turbulent flow regime in our previous work while considering periodic porous media with porosity less than 0.8. The symmetry-breaking was caused by the formation of local favorable pressure gradient regions in the pore space caused by the vortex formation and a secondary flow instability[13].

A crucial point to note in all these studies focusing on flow instabilities and bifurcations is that the porosity of the porous medium is small enough to cause the interaction of the flows around neighboring solid obstacles. Therefore, in this paper, we divide turbulent flow in porous media into three flow regimes with respect to the porosity ($\varphi$): low, intermediate, and high porosity flow regimes. At low porosity ($\varphi<0.8$), the surfaces of the solid obstacles are close to each other such that the pore spaces are equivalent to a network of interconnected channels. Recirculating microscale vortices are typically formed behind the solid obstacles at these low porosities since their formation is limited by the pore space. As a result, we only encounter the Kelvin-Helmholtz instability of the shear layers in this case. At intermediate porosity ($0.8<\varphi<0.95$), there is adequate pore space in between the solid obstacles in the intermediate porosity regime to allow the development of shedding vortices and the von Karman instability. However, the vortices interact with the neighboring solid obstacles at a downstream location. At high porosity ($\varphi>0.95$), the large pore space in between the solid obstacles results in the independence of the flow patterns surrounding the solid obstacle with respect to its neighbor. We note that vortices will still impinge on the neighboring solid obstacles downstream, but the strength of the vortices will be significantly diminished before impingement occurs.

In the present study, we are focused on turbulent flow in the intermediate porosity flow regime for cylindrical solid obstacles with circular cross-section. At intermediate porosity, the diameter of the microscale vortices that are formed behind the circular cylinders is approximately equal to the radius of the cylinder and smaller than the pore size by at least a factor of 2, resulting in complex flow phenomena and flow instabilities that we discuss in this paper. We use the Large Eddy Simulation numerical method to calculate the microscale flow distribution in the pores (section II). Numerical simulation enables us to perform our theoretical analysis with the requisite spatio-temporal resolution and three-dimensionality. We have validated our numerical methods using experimental measurements of pressure distribution in a tube bank to demonstrate the suitability of the numerical method to investigate turbulent flow in porous media. Details of the numerical model, validation, and numerical grid studies are presented in appendices A-D. We have presented our numerical results and discussed the mechanism and origin of a symmetry-breaking phenomenon occurring for turbulent flow in the intermediate porosity regime in section III.

## II. NUMERICAL METHOD

In this paper, we are investigating turbulent flow inside a periodic porous medium composed of cylindrical solid obstacles (figure 1). We model a Representative Elementary Volume (REV) of the porous medium consisting of a 4x4 in-line arrangement of solid obstacles (dimensions – $4s$ x $4s$ x $2s$). The solid obstacles are separated by a distance (center to center), $s$, which we are calling the pore size. We consider two cross-section shapes of the cylindrical solid obstacles: circular and square. We keep the solid obstacle hydraulic diameter, $d$, at a constant value of 1. We numerically simulate the microscale turbulent flow inside the pores by using Large Eddy Simulation (LES). We have shown in appendix A that the size of the REV is large enough for the convergence of the volume-averaged turbulence statistics. We apply periodic boundary conditions at all the boundaries of the REV. We sustain flow inside the REV by applying a momentum source term $g_i$. The Reynolds number of the flow ($Re$) is calculated using equation 1,

$$Re = \frac{\rho u_m d}{\mu} \tag{1}$$

where $u_m$ is the double-averaged (time and space) $x$- velocity, ρ is the density of the fluid, and μ is the dynamic viscosity of the fluid. The governing equations of momentum are non-dimensionalized by setting the fluid density, $d$, and $u_m$ equal to 1 and the dynamic viscosity equal to $1/Re$. To sustain flow at a specified $Re$, $g_i$ is adjusted iteratively at every time step to match the simulated $Re$ to the target $Re$. The transient equilibration stage where $g_i$ is adjusted is omitted from the calculation of turbulence statistics. Note that we use a different approach when we simulate the flow development during the increase in $Re$ as detailed at the beginning of section IIIB. Iterative adjustment of $g_i$ will introduce unphysical oscillations when $Re$ is increased, which is not suitable to study the temporal evolution of the flow. However, this issue is not a concern in section IIIA, where the study focuses on constant $Re$ and Reynolds-averaged flow statistics.

We use ANSYS Fluent 16.0 to solve the governing equations for LES with the Dynamic One-equation Turbulence Kinetic Energy (DOTKE) subgrid scale model by applying the finite volume method. We use bounded second order central differencing discretization[19] for the convective terms (for numerical stability) and second order central differencing discretization for the diffusive terms of the momentum equation. We solve the momentum and pressure Poisson equations in a segregated manner using the PISO algorithm[20]. We advance the simulation in time by using a second order (implicit) backward Euler method. The simulations span 400 non-dimensional time units (17,000 CPU-hours on 40 threads of Intel Xeon E5649) to converge the turbulence statistics, which corresponds to 200 flow-through cycles for a unit cell. We have confirmed that the modes of symmetry-breaking reported in the paper are not caused by unconverged turbulence statistics by verifying that the same modes are observed over 2 different simulation time intervals (supplementary material figure s1). The governing equations and details of the model implementation are provided in Appendix B. In Appendix C, we have validated our model against experimental results and demonstrated that LES with DOTKE subgrid scale modeling reproduces the experimentally measured pressure distribution on the surface of tubes in an in-line tube bank[21]. In appendix D, we have demonstrated that the grid resolution used in the present work is sufficient to resolve the turbulent scales that are relevant to the problem that we are investigating.

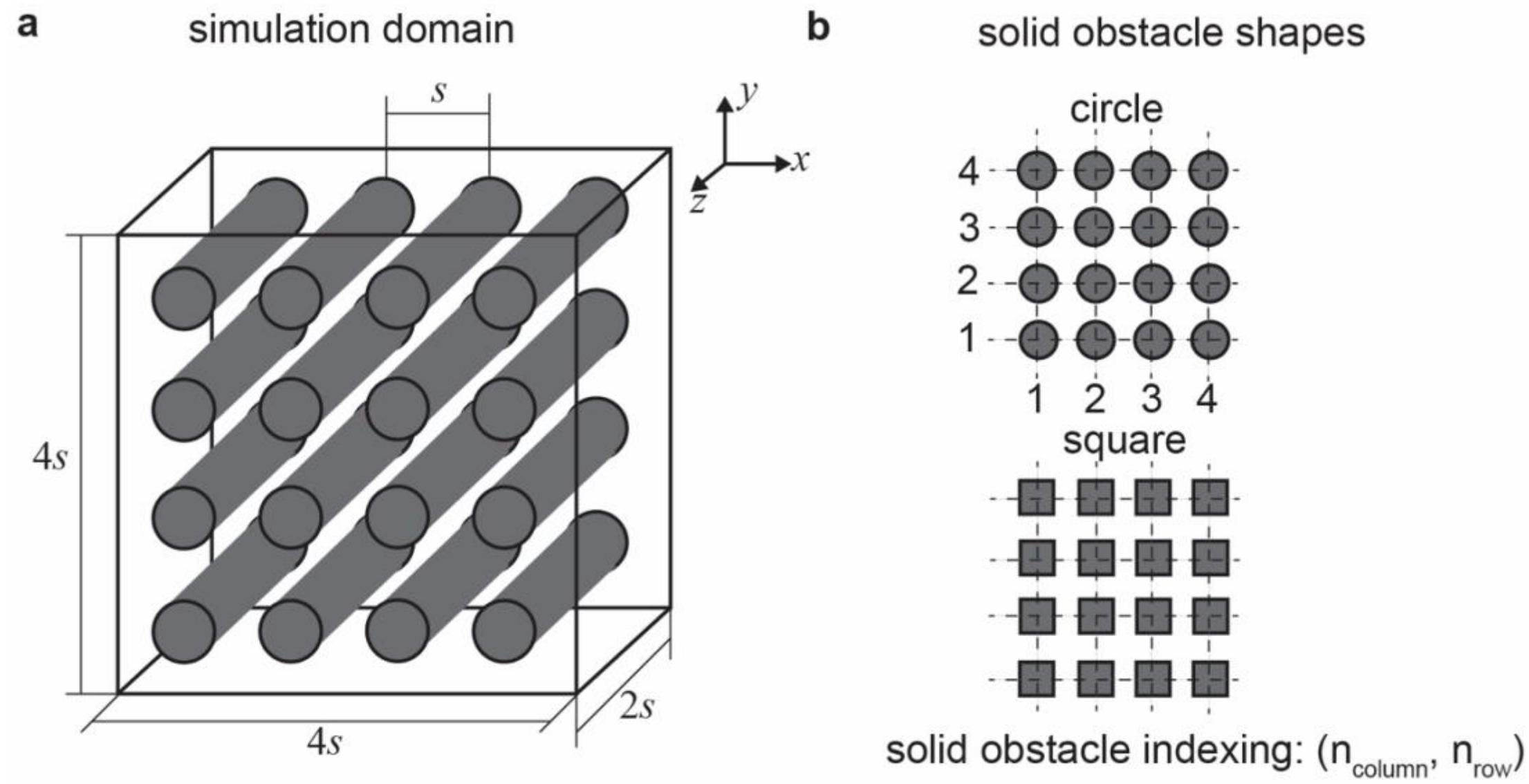

**FIG. 1.** The geometry of the computational domain used to simulate turbulent flow inside the REV of the porous medium.

## III. RESULTS AND DISCUSSION

Secondary flow instabilities and symmetry-breaking phenomena in porous media are sensitive to two key parameters: porosity and Reynolds number, and symmetry-breaking is often conditional on the smoothness of the solid obstacle surface (section I). Based on these sensitive parameters, we have structured our discussion into the following subsections:

1. A case study of the variation of porosity is used to characterize the symmetry-breaking phenomenon that occurs in the intermediate porosity flow regime.
2. A case study of the variation of Reynolds number is used to investigate the origin of the phenomenon as the flow transitions from a symmetric to asymmetric flow distribution.

### A. Characteristics of the symmetry-breaking phenomenon

Before we characterize the symmetry-breaking phenomenon in the intermediate porosity regime, we first note that turbulent flow in porous media has been observed to break symmetry in both the low ($\varphi < 0.8$)[13] and intermediate ($0.8 < \varphi < 0.95$) (this paper) porosity flow regimes. Both cases of symmetry-breaking are caused by the interaction of the flow around the solid obstacles with the neighboring solid obstacles. This is why the distance between the solid obstacle surfaces, which is determined by the porosity, is a crucial parameter in the study of flow instabilities and bifurcations occurring in porous media. Despite the commonality in the conditions that cause symmetry-breaking, its mechanism is different for the low and intermediate porosity regimes. As noted in section I, symmetry-breaking in the low porosity regime is observed for turbulent flow of sufficiently high Reynolds number ($Re \sim 500$). In this regime, a pair of shear layers are formed around the solid obstacle surface starting in the location where the flow separates (figure 2(a) and (b)). The pair of shear layers act independently of one another due to the confined pore space leading to the formation of recirculating vortices that separate them. As a result, only the Kelvin-Helmholtz instability is observed in this porosity regime at the location of the shear layers. For subcritical turbulent flow, a secondary flow instability is formed by competing inertial and pressure forces at the converging portions of the pore geometry. In this paper, we have termed this secondary flow instability as the LP (low porosity) instability. At the critical Reynolds number, the oscillation of the pressure at the stagnation point due to the LP instability causes a lateral favorable pressure gradient that results in a change in the direction of the mean macroscale flow[13]. This deviatory flow terminates at φ = 0.8 and demarcates the transition between the low and intermediate porosity flow regimes. In this paper, we want to focus our analysis on the novel discovery of symmetry-breaking in the intermediate porosity regime. Therefore, we present our case study for turbulent flow at *Re* = 300, which is subcritical for symmetry-breaking at low porosity but supercritical for intermediate porosity. We choose cylindrical solid obstacles with a circular cross-section since the smooth solid obstacle surface allows the shift in stagnation and separation points on the solid obstacle surface, unlike cylindrical solid obstacles with a square cross-section. Note that the simulation and analysis are performed in three-dimensional computational domains and the two-dimensional cross-section plots are shown here for ease of illustration and interpretation.

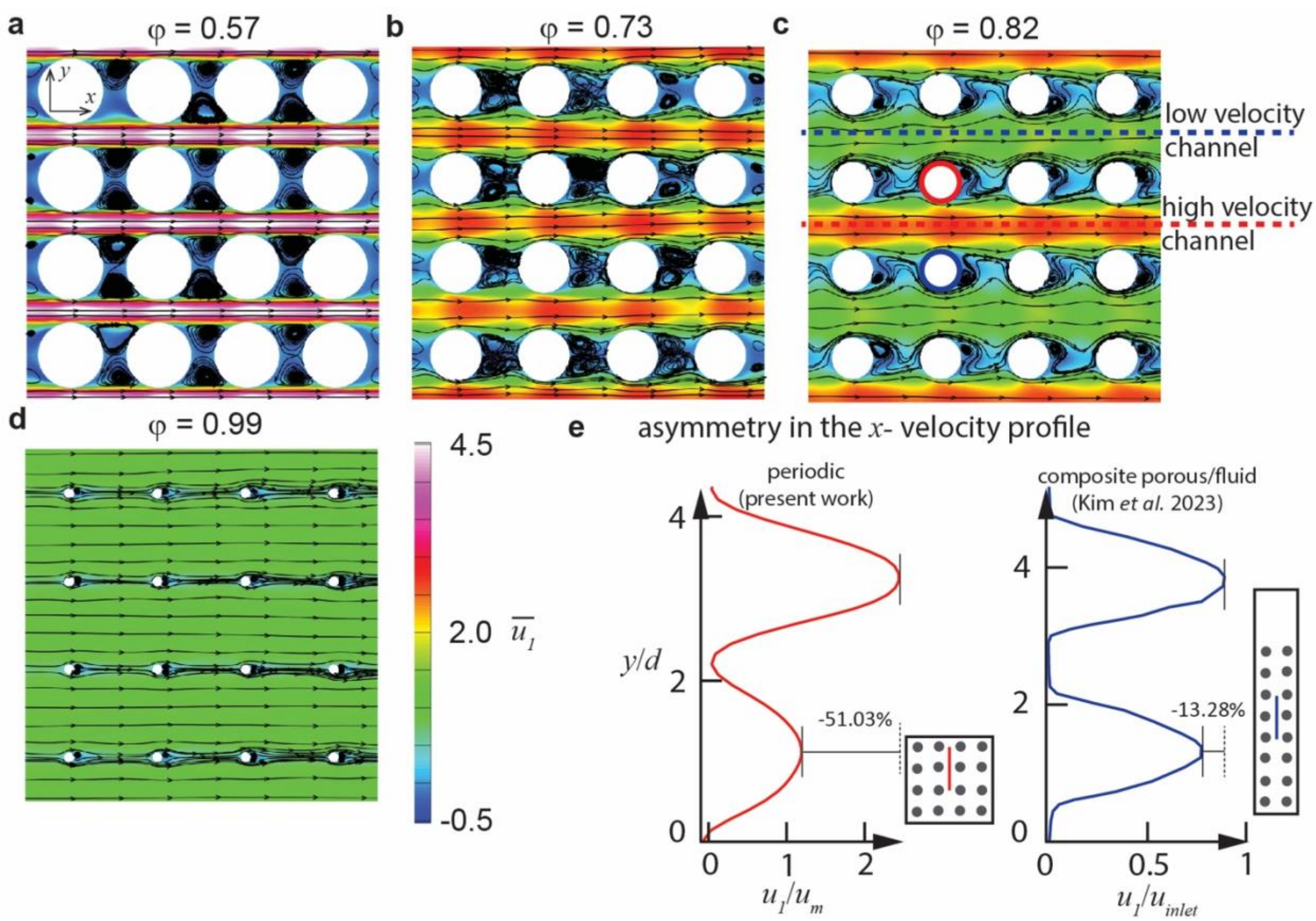

**FIG. 2.** Reynolds-averaged flow streamlines plotted for different values of porosity (a-d) show that symmetry-breaking occurs for φ = 0.82 (intermediate porosity) at *Re* = 300. The colors show the Reynolds-averaged *x*- velocity distribution. (e) *x*- velocity profiles for periodic porous and composite porous/fluid (Kim *et al.*[22]) domains in the direction perpendicular to the streamwise direction show the formation of high and low velocity channels in the pore space caused by the symmetry-breaking phenomenon. The red and blue lines in the sketches show the location of the velocity profile.

In the intermediate porosity flow regime (0.8 < φ < 0.95), a pair of shear layers are formed around the solid obstacle surface, similar to the low porosity case. The presence of an adequate pore space allows the shear layers to interact with one another resulting in vortex shedding motions similar to von Karman vortex shedding (figure 2(c)). A von Karman vortex street is not formed, however, due to the presence of the neighboring solid obstacles in the path of the vortex wake. The result is a new type of secondary flow instability that imposes a secondary oscillation on the von Karman vortex shedding. We will call this secondary flow instability as the IP (intermediate porosity) instability. The IP flow instability occurs to divert the path of the shedding vortices such that it circumnavigates the neighboring solid obstacles, which would be in the wake path otherwise. The vortices are subsequently advected into the pore space without immediately impinging on the neighboring solid obstacle surface. Thus, when the IP instability imposes a secondary oscillation on the vortex shedding process, the vortices are shed into the pore space above and below the solid obstacle in an alternating manner (figure 3). We note that neither the IP nor LP flow instabilities are observed in the high porosity flow regime due to the large separation distance between the solid obstacle surfaces (figure 2(d)).

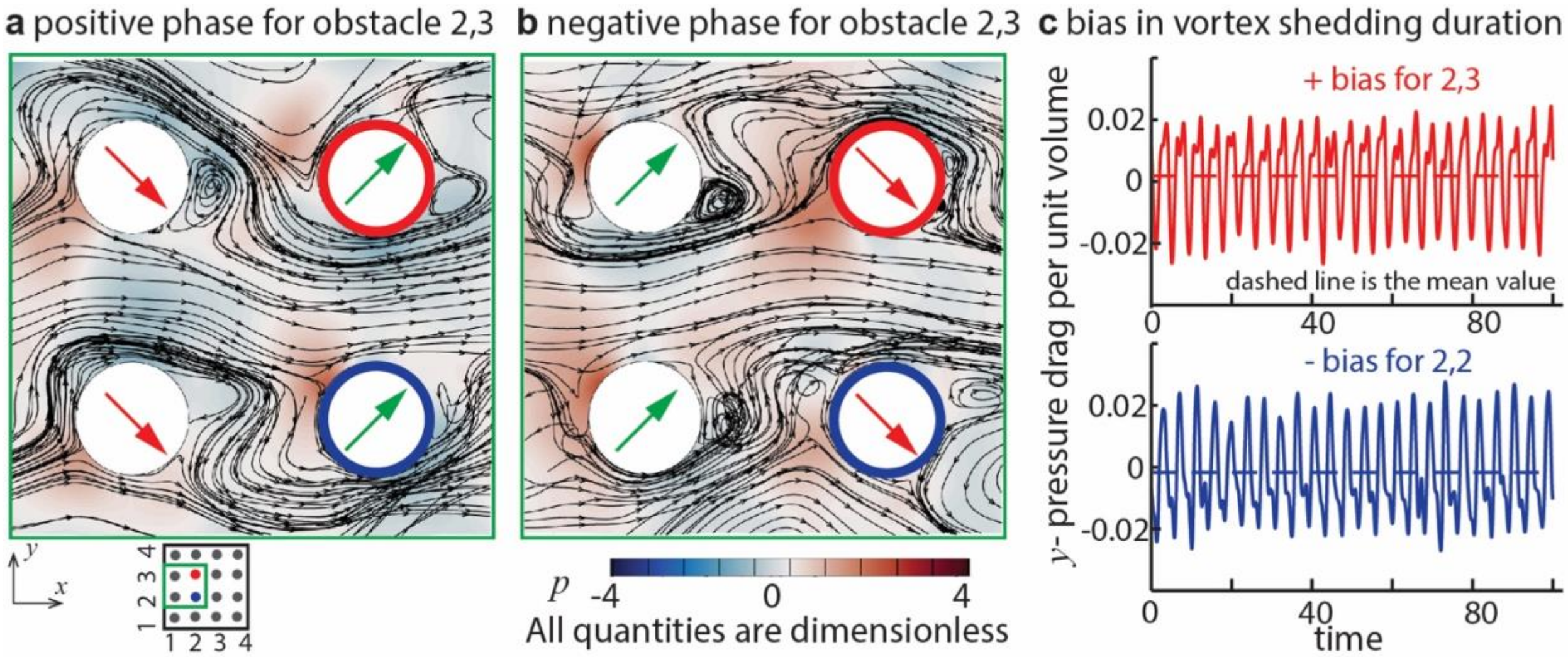

**FIG. 3.** Instantaneous flow streamlines plotted at 2 different time steps (a and b) separated by 28 non-dimensional time units show the oscillations of the IP flow instability that causes vortex shedding to alternate between the pore space above and below the solid obstacles. The colors show the instantaneous pressure distribution. The colored arrows indicate the direction of vortex shedding. (c) The time series of *y*- pressure drag force acting on the solid obstacle surface showing the bias in the duration of the vortex shedding phases that ultimately results in asymmetrical Reynolds-averaged velocity and pressure distributions. The dashed lines show the Reynolds-averaged *y*- pressure drag value for the solid obstacle. The locations of the solid obstacles (2,2 and 2,3) are shown in figure 1.

The IP flow instability introduces an oscillatory shift in the stagnation and separation points on the solid obstacle surface (figure 3). Consider the path of the vortex shedding behind the solid obstacle on the top right location in figure 3 (marked with a red circle). In figure 3(a), the vortex shedding path is tilted in the positive *y*- direction (positive phase) such that the separated flow region on the solid obstacle surface occupies the top-right quadrant. Note that a separation bubble is formed in the bottom-right quadrant, which is followed by flow reattachment on the solid obstacle surface. Correspondingly, the stagnation point is located in the bottom-left quadrant of the solid obstacle surface. When the simulation proceeds in time (figure 3(b)), the locations of the separated flow region and the stagnation point on the solid obstacle surfaces are mirrored when compared to figure 3(a) (negative phase). The streamline plots in figure 3 clearly illustrate the oscillatory behavior of the IP flow instability that introduces a shift in the separation and stagnation points along the solid obstacle surfaces. This dynamic flow behavior at the solid obstacle surface results in a large area fraction of attached flow (compared to the separated flow) when compared to the low porosity cases.

When the instantaneous flow distribution is Reynolds-averaged, the Reynolds-averaged velocity (figure 2(c)) and pressure distributions are asymmetric. This flow behavior is shown for other porosities in the intermediate regime in supplementary material figure s2. This type of symmetry-breaking that occurs in the intermediate porosity regime is different from the deviatory flow symmetry breaking that was observed at low porosities in previous work[13]. Symmetry-breaking at intermediate porosity is not evident from the instantaneous flow distribution and it does not cause the change in the mean macroscale flow direction. Asymmetry emerges in the Reynolds-averaged *x*- velocity distribution causing the formation of high and low velocity channels that alternate in the transverse direction (perpendicular to the direction of the applied pressure gradient). Here, the term channel refers to the pore space in between the solid obstacles oriented along the direction of the flow (red and blue dashed lines in figure 2(c) correspond to high and low velocity channels). Examination of the Reynolds-averaged flow streamlines in figure 2(c) reveal asymmetrical recirculation regions as well as an offset in the stagnation point with respect to the geometric plane of symmetry. In this analysis, we are referring to the geometric plane of symmetry that bisects the solid obstacle cross-section and is oriented parallel to the *xz*- plane.

These high and low velocity channels can also be observed in LES studies of a composite porous/fluid medium composed of circular tubes[22], which reported the Reynolds-averaged streamwise velocity profile in between the tubes. Although the simulation setup of Kim *et al.*[22] is different from that of the present simulation, we observe a persistent formation of channels with different centerline velocities across the span of the porous layer. We show a comparison of the simulated velocity profile for the periodic porous medium (present) and in a portion of the porous layer of a composite porous/fluid medium[22] in figure 2(e). The critical differences between the two simulation setups are as follows: (1) Kim *et al.*[22] consider a partially porous channel geometry, whereas the present work considers a fully periodic domain, and (2) the porosity of the porous medium used by Kim *et al.*[22] is 0.75, whereas the simulation uses a porosity of 0.83. A porosity of 0.75 is in the low porosity flow regime as per the definition used in the present work. However, we note that a porosity of 0.75 is close to the boundary between the low and intermediate porosity flow regimes (separated at $\varphi = 0.8$). The velocity profile for $\varphi = 0.75$ in Kim *et al.*[22] shows that the centerline channel velocity changes by 13.28% from the low velocity channel to the high velocity channel (figure 2(e)). Similarly, the velocity profile simulated in the present work for $\varphi = 0.82$ has a 51.03% change in the centerline velocity from the low velocity channel to the high velocity channel (figure 2(e)). This indicates a qualitative similarity between the results of the two simulations, as well as suggests the possibility that asymmetrical flow patterns can emerge not just in periodic porous media but also in semi-infinite porous media.

Quantitative agreement in the magnitudes of the channel centerline velocities between the two simulation studies is not feasible since the setups are different. The velocity profiles in the two studies are non-dimensionalized with respect to different characteristic velocity scales. For the simulation, the *x*- velocity is non-dimensionalized with respect to the double-averaged (space and time) *x*- velocity in the periodic REV. For Kim *et al.*[22], the *x*- velocity is non-dimensionalized with respect to the inlet velocity of the wind tunnel section. Additionally, the study reported that only a portion of the wind tunnel width is occupied by solid obstacles and the remaining portion is a clear fluid region. Therefore, with the available information about the simulation setup and velocity distribution, a direct quantitative comparison of the velocity profiles between the two studies was not possible.

In periodic porous media, the high and low velocity channels (Reynolds-averaged) develop inside the pores due to an asymmetrical vortex shedding process that occurs over time. For example, consider the high velocity channel formed in the centerline of the REV in figure 2(c) (red dashed lines). The vortex shedding behind the solid obstacle at position (2,2) (blue circle in figure 2(c)) is biased such that the negative phase of the IP instability is longer than the positive phase. Whereas the vortex shedding behind the solid obstacle at position (2,3) (red circle in figure 2(c)) is biased such that the positive phase of the IP instability is longer than the negative phase. The result is the formation of a high velocity channel in between the 2 solid obstacles where the vortex shedding process occurred over a shorter time period. Conversely, low velocity cells are formed inside the pores where the vortex shedding duration is longer.

The temporal bias in the vortex shedding process can also be observed in the *y*- pressure drag force that is acting on the surface of the solid obstacle (figure 3(c)). If the flow is symmetric, the Reynolds average of the *y*- pressure drag will be equal to zero. If the flow is asymmetric, such as in the present case, there exists a bias in the Reynolds-averaged *y*- pressure drag that acts in either the positive or negative *y*- direction depending on the offset location of the stagnation point for the Reynolds averaged flow. Note that positive bias in the *y*- pressure drag is caused because the duration of the positive phase of the vortex shedding process is greater than the duration of the negative phase (solid obstacle (2,3) in figure 3(c)). The asymmetry in the vortex shedding over time is also visible in the shape of the plot of *y*- pressure drag in figure 3(c). Positive bias in the *y*- pressure drag for the solid obstacle results in a low velocity cell in the pores above the solid obstacle. The vice-versa is observed for the solid obstacles with negative bias in the *y*- pressure drag.

The symmetry-breaking phenomenon that we have described above does not occur when the solid obstacle shape is changed from a circular to a square cross-section (supplementary material figure s2). The shift in the separation point and the oscillation of the separated flow region on the solid obstacle surface is less intense for cylindrical solid obstacles with square cross-section even though the porosity is the same as the case with circular solid obstacle cross-section. The vertices of the square geometry prescribe the location of the separation points and limit the oscillation of the vortex shedding path. As a result, vortices shed behind square solid obstacles impinge on the adjacent solid obstacles that are downstream, unlike for circular solid obstacles. The size of the vortex structures as well as the fraction of the solid obstacle surface area occupied by separated flow is larger for square solid obstacles when compared to the circular solid obstacles.

### B. Origin of the secondary flow instability and symmetry-breaking

Following the description of the characteristics and mechanism of the symmetry-breaking phenomenon described in section IIIA, we present a pointed discussion about the origin of the phenomenon in this section. We simulated the transition of the flow from a symmetric to an asymmetric flow by changing the Reynolds number of the flow. We have used cylindrical solid obstacles with circular cross-sections and kept the porosity at $\varphi = 0.82$ (same geometry as in figure 2(c)). The objective is to identify the symmetric flow conditions for this geometry and increase the Reynolds number to study the development of the secondary flow instability and subsequently, symmetry breaking. We observed that symmetric flow occurs only at laminar flow Reynolds numbers and symmetry-breaking accompanies the transition to turbulent flow. Therefore, we simulate the flow transition from $Re = 37$ (laminar) to 100 (turbulent). The asymmetric flow distribution that emerges at $Re = 100$ is sustained at higher Reynolds numbers simulated in this work ($Re =$ 300 and 1,000). Turbulence transition is also shown in supplementary material figure s3 using coherent structure visualization with the q- criterion.

As discussed in section II, we have used two approaches to sustain flow inside the periodic REV. To simulate the turbulence statistics in section IIIA, we have used a constant flow rate approach where the applied pressure gradient $g_i$ is adjusted iteratively until the desired Reynolds number is achieved. This approach is not suitable for studying the transition of the flow from a symmetric to asymmetric configuration for the following reasons: (1) we do not want to introduce unphysical oscillations in the flow solution when the Reynolds number is changed that is caused by the control system used to maintain a constant flow rate, (2) the control system dampens the flow at low Reynolds numbers ($Re = 50$) and prevents the occurrence of flow instabilities which will then only emerge at higher Reynolds numbers ($Re = 100$). This prevents us from accurately observing the transition flow behavior. Therefore, we implement the constant applied pressure gradient approach in section IIIB to study the transition from symmetric to asymmetric flow. We estimate the applied pressure gradient required to sustain a specific Reynolds number by using the constant flow rate approach and then repeat the simulation with the constant applied pressure gradient ($g_i$). If the resulting flow Reynolds number is different for the constant pressure gradient approach when compared to the constant flow rate approach, we report the exact Reynolds number that we observed and not the target Reynolds number. In all of the cases in section IIIB, we assume that the dynamic viscosity of the fluid is equal to 1/300. The results are then non-dimensionalized with respect to the velocity ($u_m$) and viscosity ($\nu$) at $Re = 300$. The applied pressure gradient ($g_i$) is increased according to the values shown in table I to simulate different flow Reynolds numbers.

Consider the laminar flow regime where the constant flow rate method is first applied to sustain a Reynolds number of 50. The resulting flow distribution simulated by the constant flow rate method is steady (figure 4(a)). When the constant applied pressure gradient method is subsequently implemented, the Reynolds number of the flow decreases to $Re = 37$ as the flow becomes unsteady (figure 4(b)). The decrease in the flow rate due to the occurrence of the IP secondary flow instability in the intermediate porosity regime can be interpreted as an increase in the drag force that is acting on the solid obstacle surface when compared to the steady case. Despite the change of unstable flow behavior on the choice of periodic flow specification algorithm, we note that the symmetry-breaking phenomenon and the secondary flow instabilities persist far beyond the transition region at higher $Re$ values of 300 and 1000 (supplementary material figure s2). The results reported in section IIIA are for $Re = 300$, far from the transition point and unaffected by the numerical approach.

**Table I**. Constant applied pressure gradient values used to sustain turbulent flow at specific Reynolds numbers and study the development of symmetry breaking.

| Reynolds number ($Re$) | Applied pressure gradient ($g_1$) |
|---|---|
| 37 | 0.00645 |
| 100 | 0.04 |
| 300 | 0.2485 |

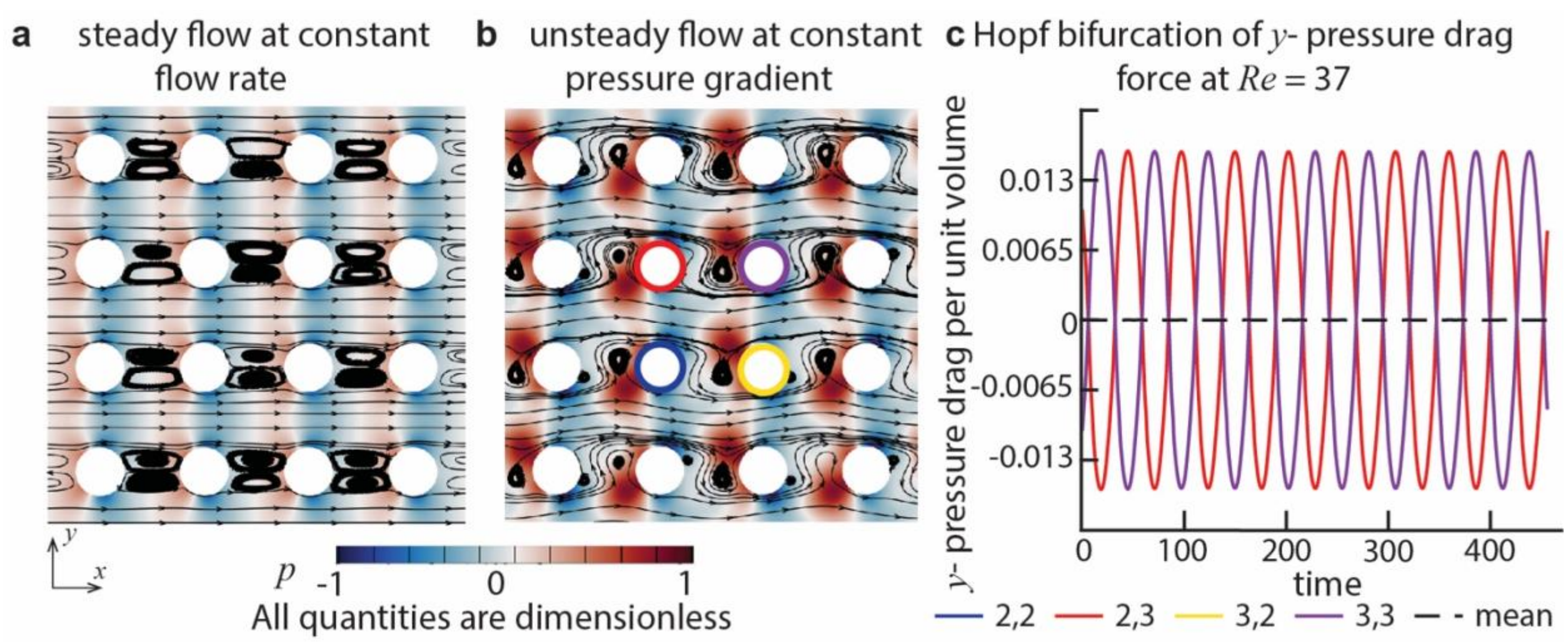

**FIG. 4.** Instantaneous flow streamlines simulated by the (a) constant flow rate method ($Re = 50$) and (b) constant pressure gradient method ($Re = 37$) for the same value of time-averaged applied pressure gradient ($g_1$). (c) A secondary flow instability develops for laminar flow at $Re = 37$ that is periodic over time.

The instantaneous flow streamlines in figure 4(b) show the oscillatory unsteady flow that develops at $Re = 37$. Even though the instantaneous flow streamlines are asymmetric in this case, the time-averaged flow solution is symmetric. The periodic nature of the oscillations observed in the $y$- pressure drag suggests that the IP secondary flow instability is a Hopf bifurcation (figure 4(c)). Similar Hopf bifurcations were observed by Agnaou *et al.*[17], who observed unsteady flow behavior for a range of solid obstacle shapes, arrangements, and porosities. When the flow transitions from a steady to an unsteady (time-periodic) state, the corresponding flow distribution changes from having a pair of stagnation points on the solid obstacle to a single stagnation point that oscillates from the top to the bottom surface of the solid obstacle (figure 4(b)). This reaffirms the "intermediate" nature of the flow features in this porosity regime caused by the boundary between the low and high porosity flow regimes. In the intermediate porosity regime, the recirculating vortices are unstable and result in the interaction of the shear layers around the solid obstacle. However, the shear layer interaction is influenced by the solid obstacle downstream leading to the formation of the periodically oscillating flow distribution. Note that the vortex oscillations behind each column of solid obstacles are out of phase by exactly one-half period (figure 4(c)). This phase difference in the vortex oscillations prevents the interference of the vortex shedding between the rows of solid obstacles in the REV.

When the Reynolds number of the flow is increased (by increasing $g_l$), the flow first transitions from the laminar to the turbulent regime. In this simulation, we have set the time at which $g_i$ is increased as $t = 0$ such that $t<0$ is the laminar flow regime and $t>0$ corresponds to the transition to the turbulent flow regime. We note that the transition from laminar to turbulent flow occurs without the presence of a transition region characterized by intermittency. We have only observed either entirely laminar or entirely turbulent flow. This can be explained by considering the process of transport of turbulence in periodic porous media where turbulence is first produced by the interaction of the microscale vortices with the mean flow around the solid obstacles. One of the key differences in the flow around a single solid obstacle and a large number of solid obstacles composing a periodic porous medium is the absence of free-stream flow conditions in the porous medium. Non-zero turbulence intensity is observed throughout the pore volume since turbulent structures produced by one solid obstacle interact with the downstream neighboring solid obstacles. As long as the Reynolds number of the flow is maintained, flow disturbances produced by the microscale vortices always result in the formation of turbulent flow structures since the incoming flow is fully turbulent due to periodicity.

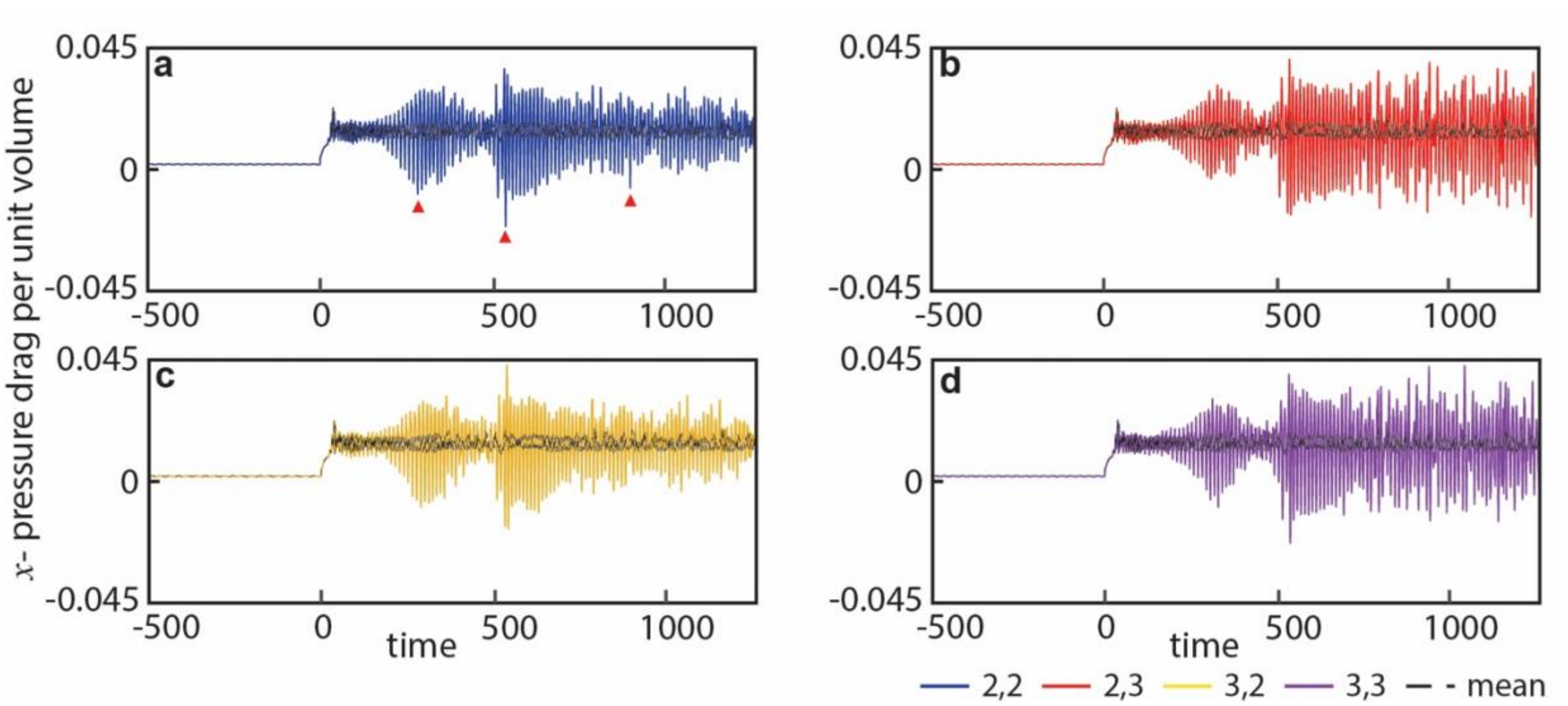

**FIG. 5.** Drag force per unit volume (non-dimensional) acting on the solid obstacle surface during the transition from laminar to turbulent flow. Panels a-d show different solid obstacles in the REV indicated by the colored lines. The time scale (non-dimensional) is adjusted such that the time at which the Reynolds number is changed is set as 0. The red triangles in panel (a) show examples of instances of negative *x*-pressure drag acting on the solid obstacle.

The resulting transition from laminar to turbulent flow involves the change in the magnitude and amplitude of oscillation of the drag forces that act on the solid obstacle surface. The periodically repeating oscillations that were observed for laminar flow at $Re = 37$ (figure 4(c)) become random oscillations for turbulent flow at $Re = 100$ (figure 5). Turbulent flow in porous media exhibits a unique characteristic where drag force oscillations are produced by both turbulent and non-turbulent phenomena[8,10]. Vortex shedding is characterized by a singular frequency (the dominant frequency) and the stochastic fluctuations of turbulent flow are superimposed on it (supplementary material figure s4). As the Reynolds number is increased, the magnitude of the pressure and shear drag force components in the *x*- direction increase in response to the increase in applied pressure gradient and the flow velocity. We also note an interesting flow behavior in the intermediate porosity regime that occurs when turbulent flow develops. The intensity of the oscillation of the *x*- pressure drag force acting on individual solid obstacles increases substantially such that the magnitude of the *x*- pressure drag force experiences negative values during some time intervals (red triangles in figure 5). This implies that the *x*- direction pressure drag force on the surface of that solid obstacle is acting in the direction that is opposite to the direction of the flow, causing a momentary propulsive force. Negative *x*- pressure drag force is experienced when the microscale vortices are impinging on the solid obstacle surface and creating low pressure regions on a substantial portion of the solid obstacle surface. Low pressure regions caused by the impingement of microscale vortices and converging pore geometry are responsible for this seemingly unphysical result – locally negative drag force for a positive applied pressure gradient. However, other solid obstacles in the REV experience positive values of drag force such that the volume-averaged *x*- pressure drag force over the REV, along with the viscous (shear) drag and inertial forces, balances the applied pressure gradient.

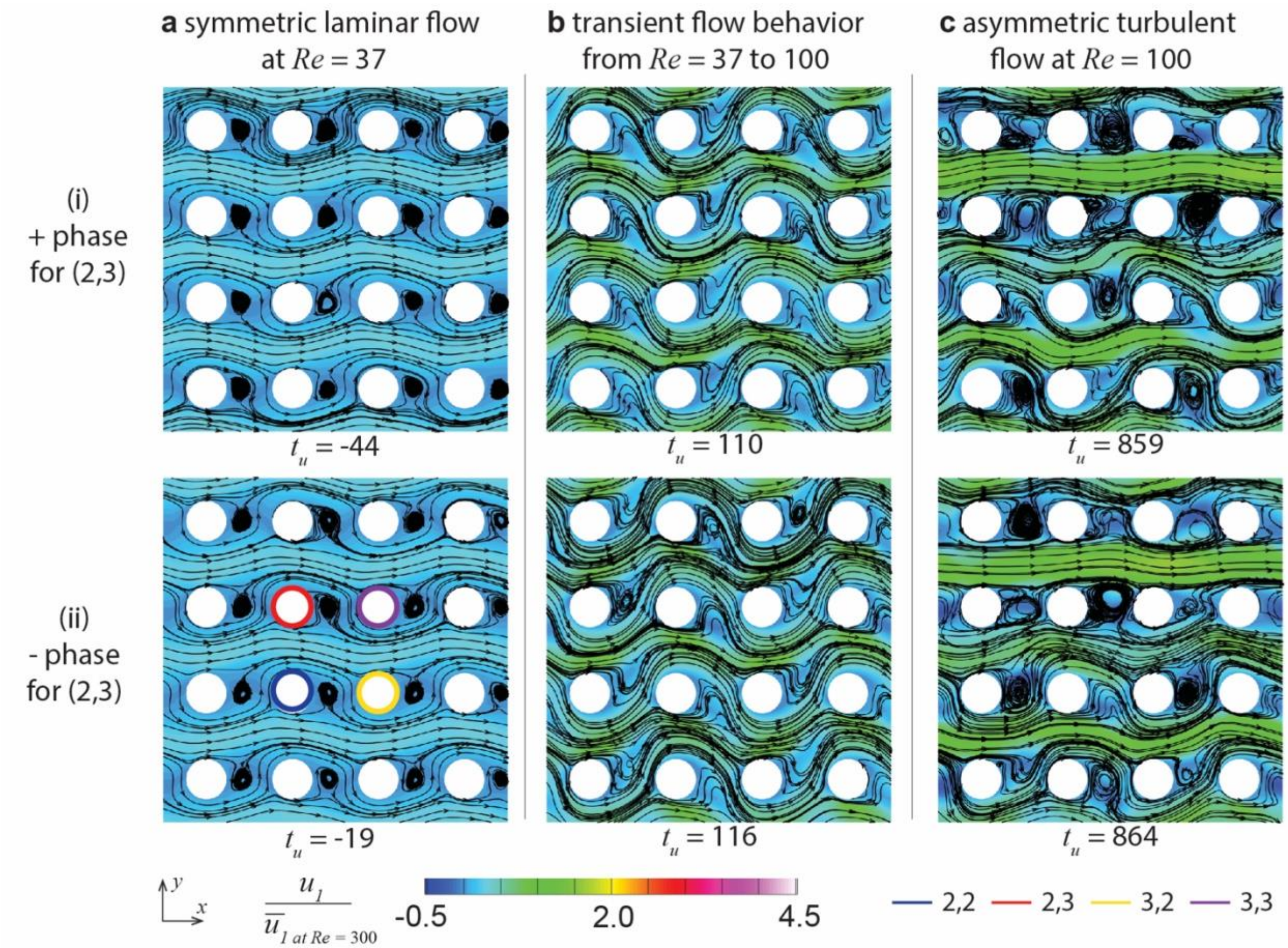

**FIG. 6.** Instantaneous flow streamlines plotted with contours of microscale *x*- velocity show the breakdown of symmetry in the flow as the flow transitions from laminar to turbulent. The three stages shown are: (a) periodic oscillations of the secondary flow instability occurring at *Re* = 37, (b) amplification of the vortex wake path oscillation when the applied pressure gradient is increased corresponding to *Re* = 100, (c) transition to turbulence at *Re* = 100 results in stochastic vortex shedding, asymmetry in the flow distribution, and the formation of high and low velocity channels.

Once the flow transitions from laminar to turbulent, asymmetry begins to develop in the microscale flow distribution inside the pores (figure 6). As discussed in section IIIA, asymmetries are observed in the pressure and velocity distributions along the *y*- direction. Therefore, we complement the analysis of the instantaneous flow streamlines (figure 6) with the time evolution of the *y*- pressure drag force that is acting on the individual solid obstacle surfaces (figure 7). For laminar flow ($t < 0$), the periodic unsteady flow that is caused by the Hopf bifurcation is observed in the flow streamlines (figure 6(a)). Correspondingly, the oscillations of the *y*- pressure drag force are in phase for all the solid obstacles belonging to a single column in the REV ((2,2) and (2,3) in figure 7(a)), whereas the instability is out of phase by one-half period for solid obstacles belonging to the adjacent column in the REV ((2,2) and (3,2) in figure 7(b)). This synchronization in the flow instability across the solid obstacles in the REV is possible because the flow is laminar. When the Reynolds number is increased at $t = 0$, the amplitude of spatial oscillations of the flow streamlines around the solid obstacles increases (figure 6(b)), as well as the intensity of the temporal oscillations of the *y*- pressure drag force (figure 7). When the amplitude of spatial oscillations of the flow streamlines increases, it leads to the interference of the vortex wake path behind each solid obstacle with the solid obstacles in the adjacent rows of the REV. The synchronization of the flow instability during laminar flow prevents the collision of vortex wake paths as observed in figures 6(b) because of the periodicity of vortex shedding.

Randomness in the flow develops between $t = 200$ to 400 during the transition to turbulent flow. The initially orderly oscillations of the flow instability break down into stochastic flow behavior. Consequently, the phase difference in the drag force instability between neighboring solid obstacles deviates from zero and one-half period (column and row) observed at $t = 0$. Stochastic variation in the phase difference for all the solid obstacles (both columns and rows in the REV) develops over the time period from $t = 200$ to 400 (figure 7). Following the development of stochasticity due to turbulence, the amplitudes of oscillation of the *y*- pressure drag force are similar for solid obstacles that are in the same row, whereas the amplitudes are different for solid obstacles in the same column. This observation coincides with the observation of high and low velocity channels forming between the solid obstacle rows (figure 6(c)), which leads to the asymmetric flow distribution as described in section IIIA. The formation of asymmetry prevents the collision of vortex wake paths between the rows of solid obstacles since the bias in the vortex shedding regulates vortex advection. Note that the mode of symmetry-breaking in the present case (low-high-low-high) is different from the mode of symmetry breaking in figure 2(c) (high-low-high-low) even though the porosity is equal to 0.82 in both cases. The formation of the stochastic phase difference and subsequent development of asymmetrical vortex shedding determine the mode of symmetry breaking and the locations of the high and low velocity channels in the REV.

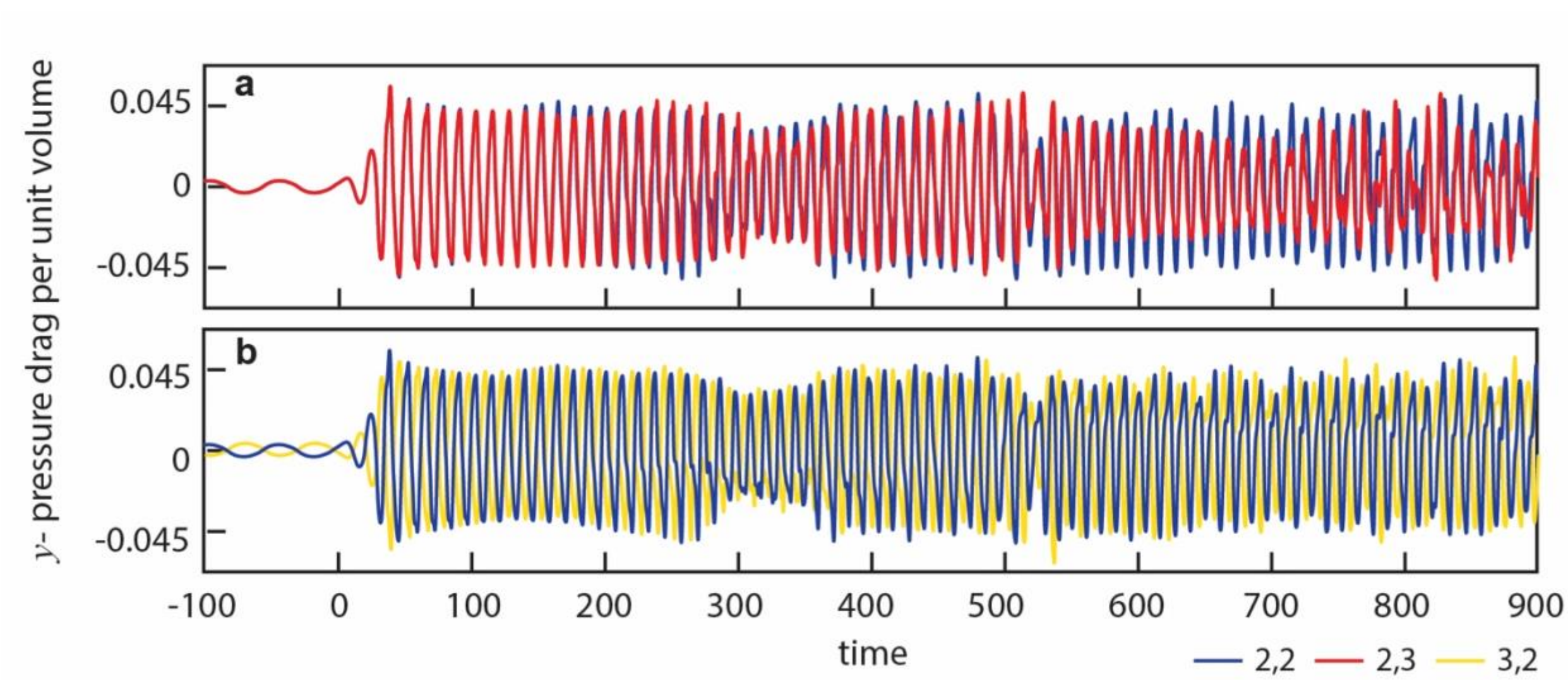

**FIG. 7.** *y*- pressure drag force acting on the solid obstacle surface of obstacles (2,2), (2,3), and (3,2) shows the transition from one-half period phase difference (laminar) to stochastic phase difference (turbulent) in the vortex shedding behind columns of solid obstacles in the REV.

## IV. SUMMARY

In this paper, we have simulated turbulent flow in periodic porous media in the intermediate porosity regime to analyze the origin and mechanism of flow symmetry breaking. The porous medium is composed of an in-line arrangement of cylindrical solid obstacles with circular cross-section. We characterize three flow regimes for turbulent flow in porous media based on the behavior of the microscale vortex structures at the different values of porosity:

1. Low porosity regime ($\varphi<0.8$) is characterized by the formation of the Kelvin-Helmholtz instability and the formation of recirculating vortices in between the solid obstacles.
2. Intermediate porosity regime ($0.8<\varphi<0.95$) is characterized by the formation of the von Karman vortex shedding instability. Vortex shedding is influenced by the presence of solid obstacles at a downstream location.
3. High porosity regime ($\varphi>0.95$) is characterized by the formation of the von Karman vortex shedding instability, which is virtually independent of the neighboring solid obstacles.

We note that the range of porosities characterizing the low, intermediate, and high porosity regimes is specific to cylindrical solid obstacles with circular cross-section and may change for different solid obstacle shapes.

In the intermediate porosity flow regime, we report the occurrence of the intermediate porosity (IP) secondary flow instability and a symmetry-breaking phenomenon for cylindrical solid obstacles with circular cross-section, which causes asymmetry in the microscale velocity and pressure distributions inside the pore volume. The IP secondary flow instability is caused by the interaction of the shear layers that are formed around the solid obstacle surface during the formation of the microscale vortices with the solid obstacles at a downstream location. This causes the oscillation of the vortex wake behind the solid obstacle, advecting vortex structures into the pore space rather than impinging on the neighboring solid obstacle surface. The oscillation of the vortex wake is accompanied by a shift in the separation point on the solid obstacle surface which decreases the size of the vortex recirculation. Thus, the resulting flow pattern surrounding the solid obstacle surface includes more surface area covered by the attached flow.

The symmetry of the Reynolds-averaged velocity and pressure distributions is broken because the vortex wake oscillations caused by the IP secondary flow instability are not symmetric with respect to time. To analyze this, we divided the oscillations of the flow instability into positive and negative phases based on the direction of the pressure drag force that is acting on the solid obstacle surface. We noted that either the positive phase of the oscillation spans a greater duration than the negative phase, or vice versa. When the instantaneous flow distribution is averaged over time, the Reynolds-averaged velocity and pressure distributions are asymmetric. The asymmetry is more prominent for the velocity distribution leading to the formation of high and low velocity channels in the pore space when compared to the pressure distribution. These high and low velocity channels are oriented along the direction of the volume-averaged flow velocity. The velocity channels are alternating between high and low velocity in the transverse direction of the flow. Low velocity channels are formed in the pore spaces where the vortices are preferentially advected due to a temporal bias in the vortex shedding phases. High velocity channels are formed in the complementary pore spaces to the low velocity channels since there is less recirculating motion in these pore spaces when the flow is averaged over time. Asymmetry emerges in the pressure distribution as a shift in the location of the stagnation points on the solid obstacle surface. We noted that the oscillatory vortex wake is limited for cylindrical solid obstacles with square cross-section by the vertices of the square geometry. These vertices prevent the shift in the separation point. Therefore, symmetry-breaking is not observed for cylindrical solid obstacles with a square cross-section.

We studied the transition of the flow from symmetrical (laminar) to asymmetrical (turbulent) distribution and observed that symmetry-breaking is accompanied by the transition to turbulence. The IP secondary flow instability develops in the laminar flow regime resulting in unsteady harmonic oscillations of the flow streamlines around the solid obstacles and the pressure drag force acting on the solid obstacle surface. We observed that solid obstacles in the same column of the REV have zero phase difference in the pressure drag force. Whereas solid obstacles in the adjacent columns of the REV have exactly a one-half period phase difference in the pressure drag force. When the flow transitions to turbulence, (1) the amplitudes of the spatial oscillations in the flow streamlines and the temporal oscillations of the *y*- pressure drag force increase, and (2) stochastic phase difference in the *y*- pressure drag force emerges for every solid obstacle regardless of the row or column in the REV where it is located. Loss of synchroneity in the vortex formation behind the solid obstacles of the REV coincides with the occurrence of symmetry-breaking in the flow. This observation suggests that symmetry-breaking of turbulent flow in porous media in the intermediate porosity regime is a regulating mechanism for the vortex shedding process with secondary flow instability. Before symmetry-breaking occurs, the amplitude of oscillation of the vortex wake path is large enough to interfere with the flow around the neighboring solid obstacles in the transverse direction to the volume-averaged flow velocity. If the high amplitude of vortex wake oscillations is combined with the stochasticity of turbulence, there is a possibility of interference of the vortex motions behind neighboring solid obstacles in the REV. However, we observe that this scenario does not occur since symmetry-breaking develops. The formation of the velocity channels, where vortices are preferentially shed into the low velocity channel, regulates the vortex shedding process such that vortices do not collide.

## SUPPLEMENTARY MATERIAL

This article includes online supplementary material containing figures that provide additional results at the end of the document.

## ACKNOWLEDGEMENTS

The authors acknowledge the computing resources provided by North Carolina State University High Performance Computing Services Core Facility (RRID:SCR_022168). AVK acknowledges the support of the Alexander von Humboldt Foundation through the Humboldt Research Award. This research was funded by the National Science Foundation award CBET-2042834

## AUTHOR DECLARATIONS

### Conflict of interest

The authors report no conflict of interest.

### Author Contributions

**Vishal Srikanth:** conceptualization, investigation, and writing - review and editing (equal); Software, formal analysis, and writing - original draft (lead). **Andrey V. Kuznetsov:** conceptualization, investigation, and writing - review and editing (equal); project administration and funding acquisition (lead); formal analysis (supporting).

## DATA AVAILABILITY

The data that support the findings of this study are available within the article.

### Appendix A: Adequacy of the REV size

Since we are using an REV with periodic boundary conditions, we must ensure that the simulation captures the largest turbulent eddies. Numerical investigations using smaller[23,24] and larger[8,10] periodic REVs than the ones used in the present work can be found in the literature. Direct numerical simulation studies using large REVs (for example, 10x10), as well as experimental studies[9,25], show that the size of the largest turbulent structures in porous media is comparable to the pore size. In addition to resolving large scale eddies, the size of the REV must be large enough for the volume-averaged turbulence statistics to converge. To investigate this, we have chosen a representative simulation case of flow at $Re$ = 1,000 through an REV with cylindrical solid obstacles with circular cross-section such that the porosity is 0.8. We vary the size of the REV from $1s$ to $5s$ in increments of $1s$. We calculate the double-averaged applied pressure gradient, which is representative of the total drag force acting on the solid obstacles, and the volume-averaged turbulence kinetic energy (TKE) to show convergence of second-order turbulence statistics.

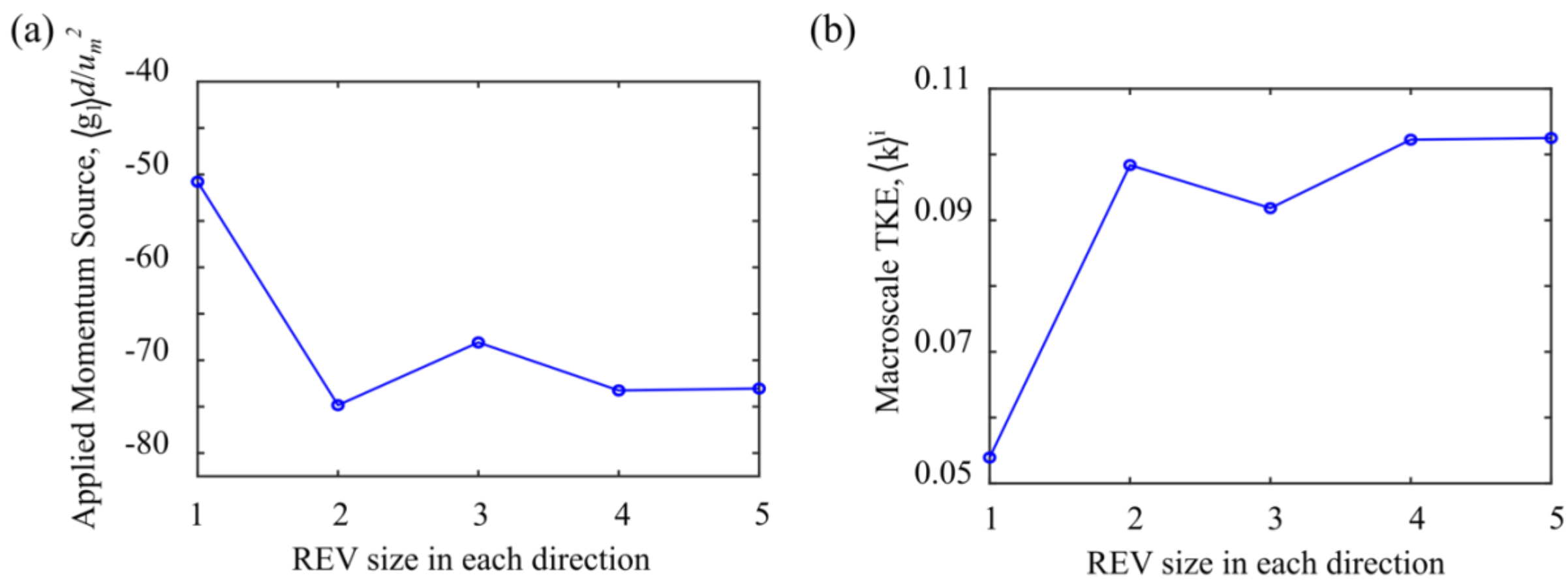

**FIG. 8.** Variation of the (a) applied pressure gradient and (b) volume-averaged TKE with respect to the size of the REV (in units of *s*).

Both the applied pressure gradient and the volume-averaged TKE converge after the REV size of 4*s* (figure 8). When the REV size is increased from 4*s* to 5*s*, the applied pressure gradient and volume-averaged TKE change by 0.4% and 0.25%, respectively. This leads us to conclude that an REV size of 4*s* is adequate to simulate the turbulent flow in the periodic REVs used in the present work.

**Appendix B: Governing equations of the Large Eddy Simulation model**

We have used the spatially filtered Navier-Stokes equations for LES with implicit filtering (the finite volume grid filters the flow field below the size of the grid cells). We use the Dynamic One-equation Turbulence Kinetic Energy (DOTKE) subgrid scale model to account for the turbulence kinetic energy (TKE) of the eddies smaller than the grid size. The governing equations of the model implemented in ANSYS Fluent 16.0 are shown in equations B.1-B.8:

$$\frac{\partial \widetilde{u_j}}{\partial x_j} = 0 \tag{B.1}$$

$$\frac{\partial \rho \widetilde{u_i}}{\partial t} + \frac{\partial \rho \widetilde{u_i}\widetilde{u_j}}{\partial x_j} = -\frac{\partial \tilde{p}}{\partial x_i} + \frac{\partial}{\partial x_j}\left[\left(\mu + \mu_{T,SGS}\right)\left(\frac{\partial \widetilde{u_i}}{\partial x_j} + \frac{\partial \widetilde{u_j}}{\partial x_i}\right)\right] + \rho g_i \tag{B.2}$$

$$\frac{\partial k_{SGS}}{\partial t} + \frac{\partial(\widetilde{u_j} k_{SGS})}{\partial x_j} = \left[C_k k_{SGS}^{\frac{1}{2}} \Delta \left(\frac{\partial \widetilde{u_i}}{\partial x_j} + \frac{\partial \widetilde{u_j}}{\partial x_i}\right)\right]\frac{\partial \widetilde{u_i}}{\partial x_j} - C_\varepsilon \frac{k_{SGS}^{\frac{3}{2}}}{\Delta} + \frac{\partial}{\partial x_j}\left(\mu_{T,SGS}\frac{\partial k_{SGS}}{\partial x_j}\right) \tag{B.3}$$

$$\mu_{T,SGS} = C_k k_{SGS}^{\frac{1}{2}} \Delta \tag{B.4}$$

$$\tau_{ij} = -2C_k {k_{SGS}}^{\frac{1}{2}} \Delta \widetilde{S_{ij}} + \frac{2}{3}\delta_{ij} k_{SGS}, L_{ij} = -2C_k {k_{test}}^{\frac{1}{2}} \hat{\Delta} \widehat{\widetilde{S_{ij}}} + \frac{1}{3}\delta_{ij} L_{kk} \tag{B.6}$$

$$C_k = \frac{1}{2}\frac{L_{ij}\sigma_{ij}}{\sigma_{ij}\sigma_{ij}}, \sigma_{ij} = -\hat{\Delta} {k_{test}}^{\frac{1}{2}} \widehat{\widetilde{S_{ij}}}, k_{test} = \frac{1}{2}\left(\widehat{\widetilde{u_k}\widetilde{u_k}} - \widehat{\widetilde{u_k}}\widehat{\widetilde{u_k}}\right) \tag{B.7}$$

$$C_\varepsilon = \frac{\widehat{(\partial \widetilde{u_i}/\partial x_j)(\partial \widetilde{u_i}/\partial x_j)} - (\partial \widehat{\widetilde{u_i}}/\partial x_j)(\partial \widehat{\widetilde{u_i}}/\partial x_j)}{\left((\mu + \mu_{T,SGS})\hat{\Delta}\right)^{-1} {k_{test}}^{\frac{3}{2}}} \tag{B.8}$$

where $u$ is the microscale velocity, $p$ is the microscale pressure, and $k_{SGS}$ is the subgrid TKE. The tilde denotes spatial filtering with the subgrid scale filter length Δ calculated as the cube root of the cell volume.

We estimate the model constants $C_k$ and $C_\varepsilon$ using a dynamic procedure[26] where the grid scale velocity field is filtered to a second, test scale velocity field. The test filter length $\hat{\Delta}$ is set equal to $2\Delta$ [27]. The circumflex denotes test scale spatial filtering. We limit the value of $C_k$ by $-\mu/\left({k_{SGS}}^{1/2}\Delta\right)$ to avoid a negative total viscosity ($\mu + \mu_{T,SGS}$). The dynamic subgrid scale formulation theoretically yields $\mu_{T,SGS} = 0$ at laminar flows allowing its application in section IIIB near the transition region.

**Appendix C: Validation of the numerical model**

The LES model is validated against experimental measurements of surface pressure distribution for turbulent flow through in-line tube banks[21]. This experimental geometric setup is similar to the REV geometry that we have used in the present work. In the experiment, a 7x4 tube bank is placed in the test section of a wind tunnel occupying the entire width of the test section. We have chosen the pressure distribution on the 4th and 7th tube columns for the case setup with $s/d = 1.6$ and $Re = 41{,}000$ for validation. This corresponds to a porosity of 0.7, which falls in the low porosity flow regime. Therefore, we do not expect to see the high porosity symmetry-breaking observed in the present work. The Reynolds number of the experiment is 40-100 times higher than the Reynolds numbers used in the present work. We note that it is beneficial to use a larger Reynolds number for validation to demonstrate the LES model accuracy when a substantial portion of the TKE is estimated by the subgrid model. For the Reynolds numbers used in the present work, the performance will only improve when compared to the validation case.

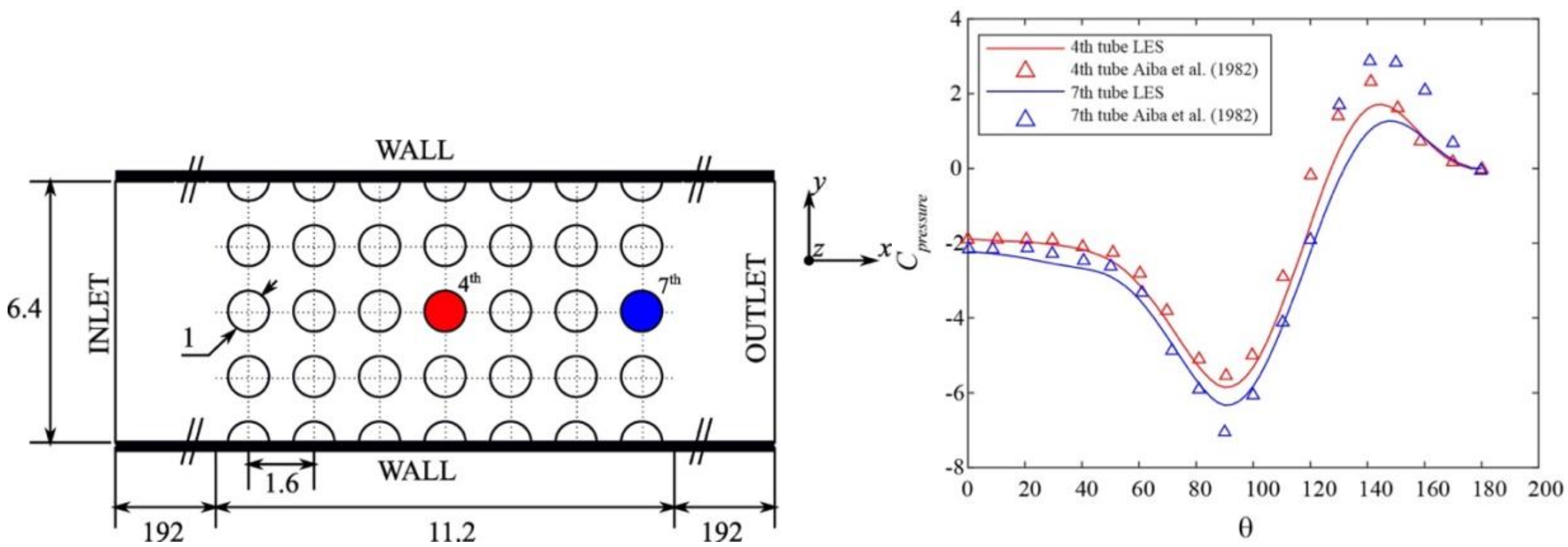

**FIG. 9.** (a) The computation domain that is used for the validation case to reproduce the experimental results of Aiba *et al.*[21]. (b) The distribution of the coefficient of pressure on the surfaces of tubes 4 (blue) and 7 (red) in the center row of the tube bank for the LES (solid line) and the experiment (triangle symbols).

We have made several assumptions for quantities that are not reproducible from the experiment. The geometry of the validation case is shown in figure 9. We assume that the inlet of the wind tunnel section is sufficiently long (192*d*) to fully develop the boundary layer and the outlet of the wind tunnel is sufficiently long (192*d*) to assume atmospheric conditions at the exit. We apply periodic boundary conditions along the spanwise direction of the tubes. The resulting coefficient of pressure ($c_{pressure}$) distribution from the simulation closely matches the experimental measurements (figure 9). The qualitative features such as the formation of a pair of stagnation points and flow separation are simulated at the exact locations they are observed in the experiment. The quantitative agreement in the $c_{pressure}$ magnitude is excellent on most of the solid obstacle surface with model error emerging at the stagnation point. We note that there are considerable differences between the simulation and experimental setup, which can prevent an exact agreement in the $c_{pressure}$. The assumptions of periodicity in the spanwise direction and arbitrary specification of the wind tunnel test section entry and exit lengths can contribute to the simulation error. However, these results suggest that the numerical method used in the present work reproduces the flow features that are observed in porous media (tube banks) in experimental work. The numerical accuracy will be better for cases in section III when compared to the validation case due to the higher resolution grids and the lower Reynolds number used in the present work.

**Appendix D: Adequacy of the grid resolution**

In the present study, our goal is to investigate the large scale turbulent vortex structures, large scale flow instabilities, and the behavior of the macroscale flow variables. We use LES for our study by assuming that the fine scale eddies have a negligible influence on the flow behavior of the large scale eddies. We show that the grid resolution used in the present work is adequate to complete our objectives. For the grid resolution study, we consider turbulent flow at *Re* = 1,000 for 4 values of porosity 0.5, 0.61, 0.72, and 0.8. We kept the near-wall grid resolution at a constant value of 0.001*s*, which we have determined to be adequate to ensure that the non-dimensional near-wall grid height ($\Delta y^+$) is consistently less than 1 (supplementary material figure S5). We tested three values of the maximum grid size ($\Delta x/s$): 0.01, 0.02, and 0.03.

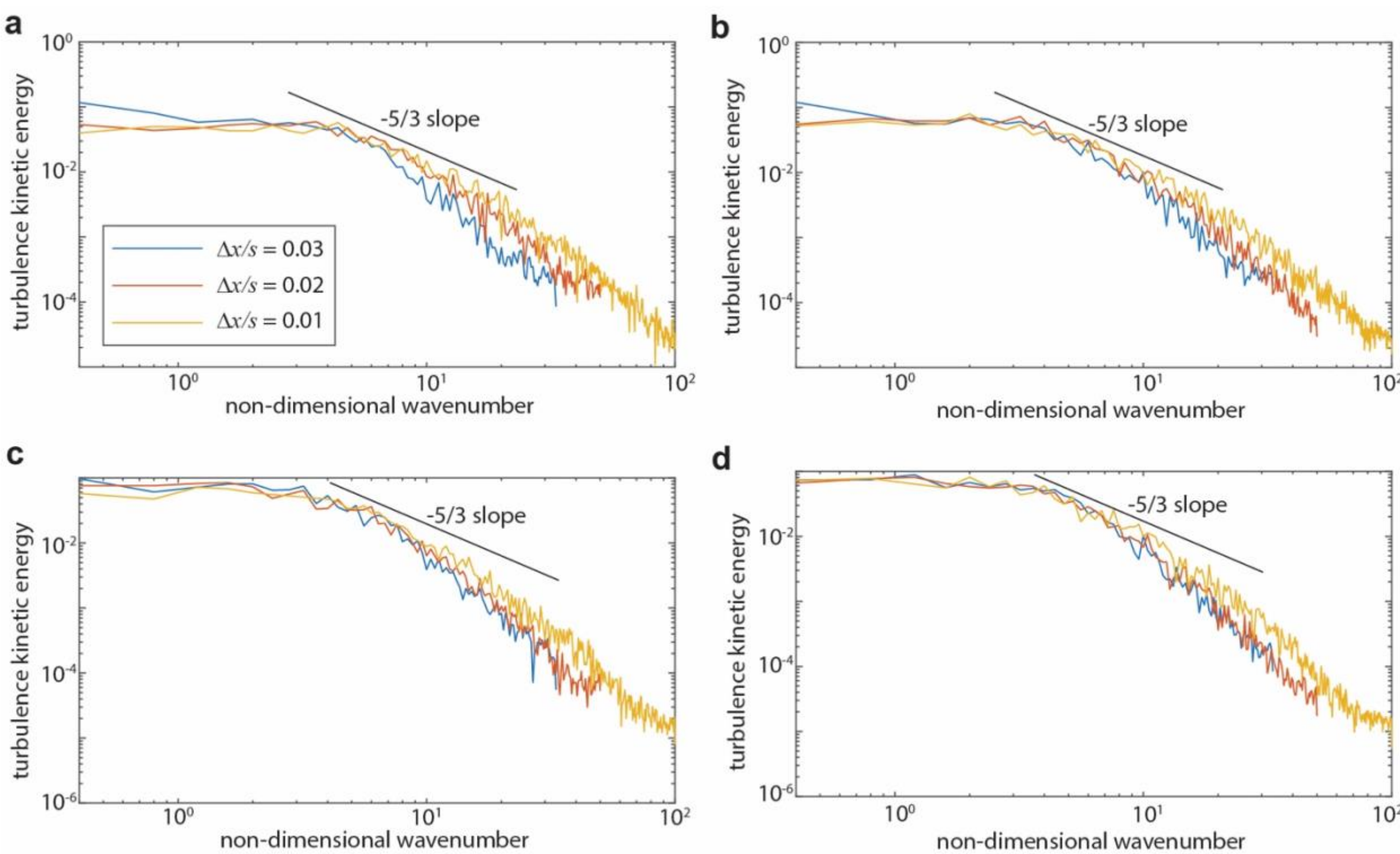

**FIG. 10.** Wavenumber spectra of turbulence kinetic energy at different grid resolutions ($\Delta x/s$) for porosities (a) φ = 0.5, (b) φ = 0.6, (c) φ = 0.72, (d) φ = 0.8. The wavenumber is scaled with respect to *s*.

**Table II.** Percentage change in the total drag force magnitude due to grid refinement

| Change in the grid resolution | φ = 0.5 | φ = 0.8 |
|---|---|---|
| $\Delta x/s$ = 0.03 to 0.02 | 3.2% | 2.2% |
| $\Delta x/s$ = 0.02 to 0.01 | 0.6% | 1.2% |

The TKE spectra (figure 10) are virtually identical for the large scale eddies at the different grid resolutions showing that the increase in the grid resolution has a negligible effect on the large scale turbulent motions. Differences in the TKE spectra emerge for the dissipative eddies after the slope of the spectrum plot decreases below -5/3. The dissipative eddies have low spectral intensity of TKE with the smallest eddies having 3 orders of magnitude less intensity when compared to the largest eddies. As a result, we observe only a small improvement (for $\Delta x/s$ = 0.02 to 0.01 in table II) in the calculation of the total drag force. Therefore, we use a grid resolution of $\Delta x/s$ = 0.02 for the simulations in section III. We note that resolving fine scale eddies will not provide new information when studying the large scale turbulent vortices and flow instabilities. We choose to model the contribution of the fine scale eddies with the use of a subgrid model to decrease the computational cost of the simulations.

**Online Supplementary Material**

Symmetry-breaking of turbulent flow due to asymmetric vortex shedding in periodic porous media

Vishal Srikanth and Andrey V. Kuznetsov

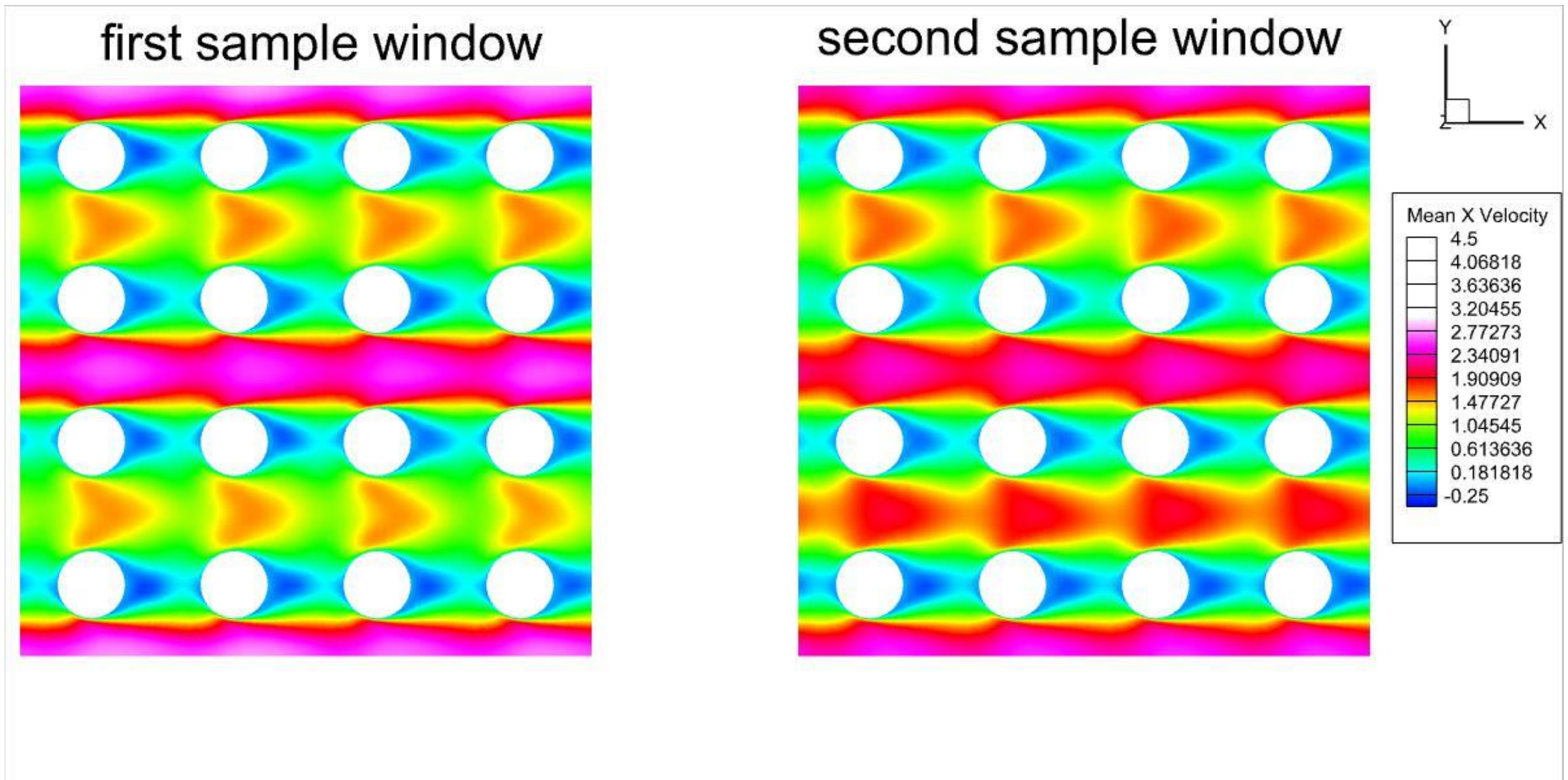

FIG. S1. The same symmetry breaking mode is recovered over two sampling windows for φ = 0.82 and *Re* = 300, each spanning over 100 non-dimensional time units.

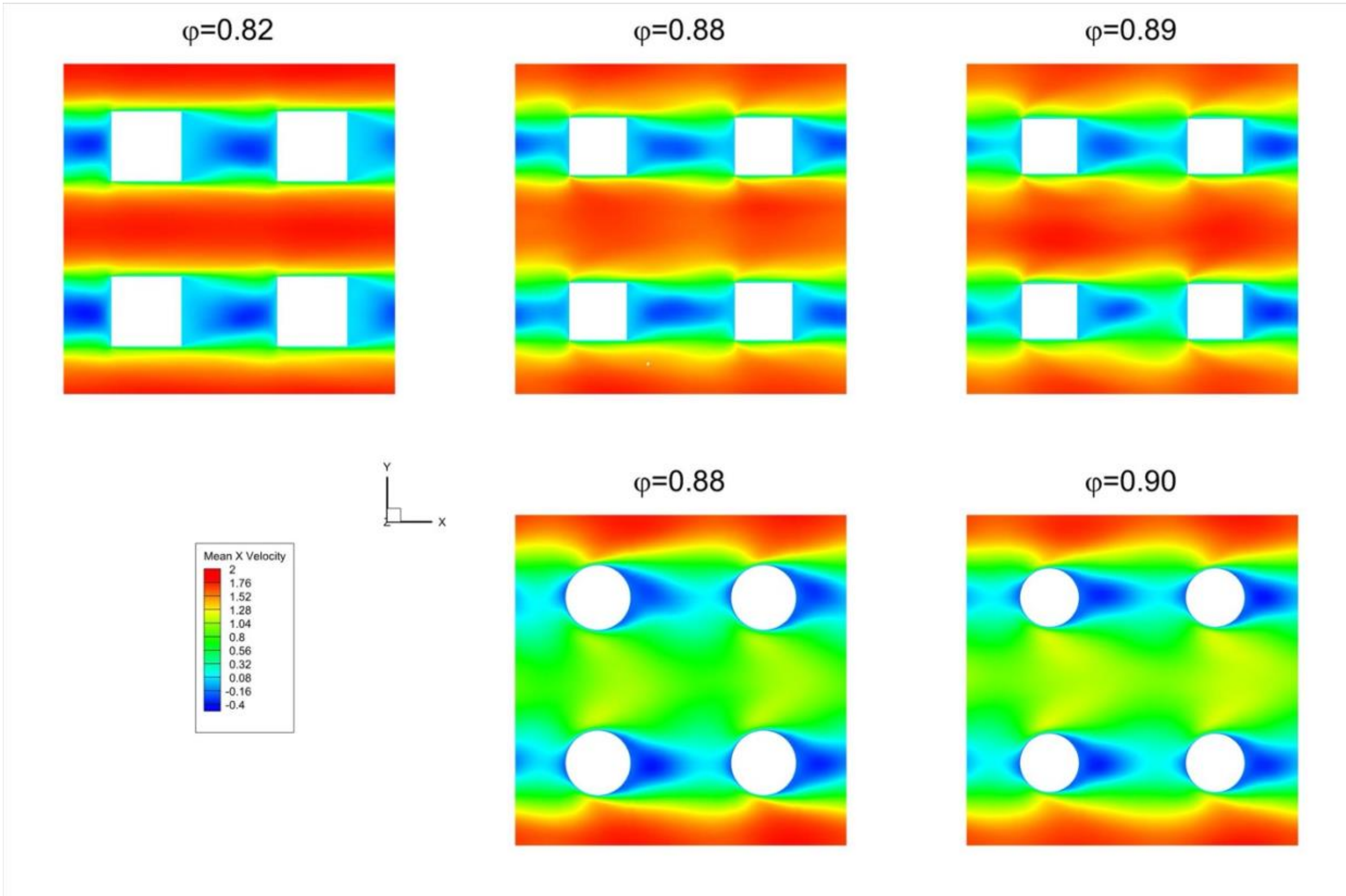

**FIG. S2.** Reynolds averaged *x-* velocity distributions at *Re* = 1000 showing that symmetry-breaking does not occur for square cylinder solid obstacles. Additionally, this data demonstrates that symmetry-breaking occurs for circular cylinder solid obstacles at higher *Re* and different porosity values within the intermediate regime.

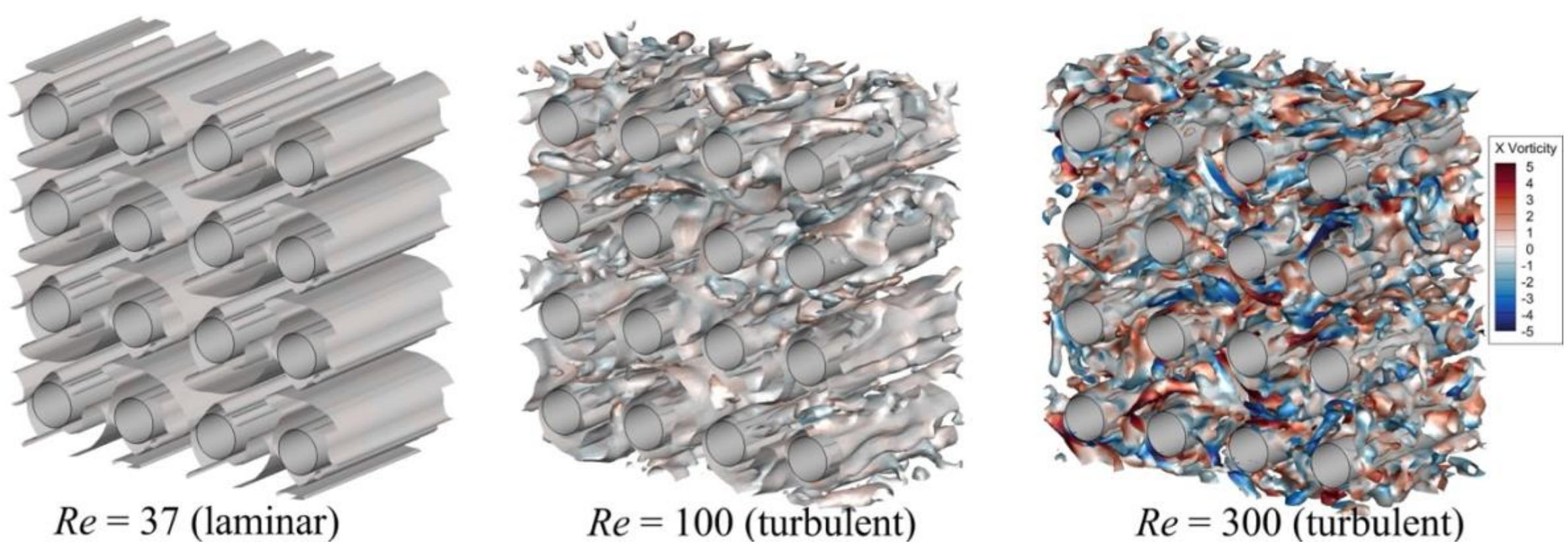

**FIG. S3.** Iso-surfaces of q- criterion are used to visualize vortical flow structures at *Re* = 37, 100, 300. The iso-surfaces are colored by contours of non-dimensional *x-* vorticity. The emergence of randomly shaped 3D coherent structures and the non-zero *x-* vorticity magnitude proves that the flow is turbulent at *Re* = 100 and 300.

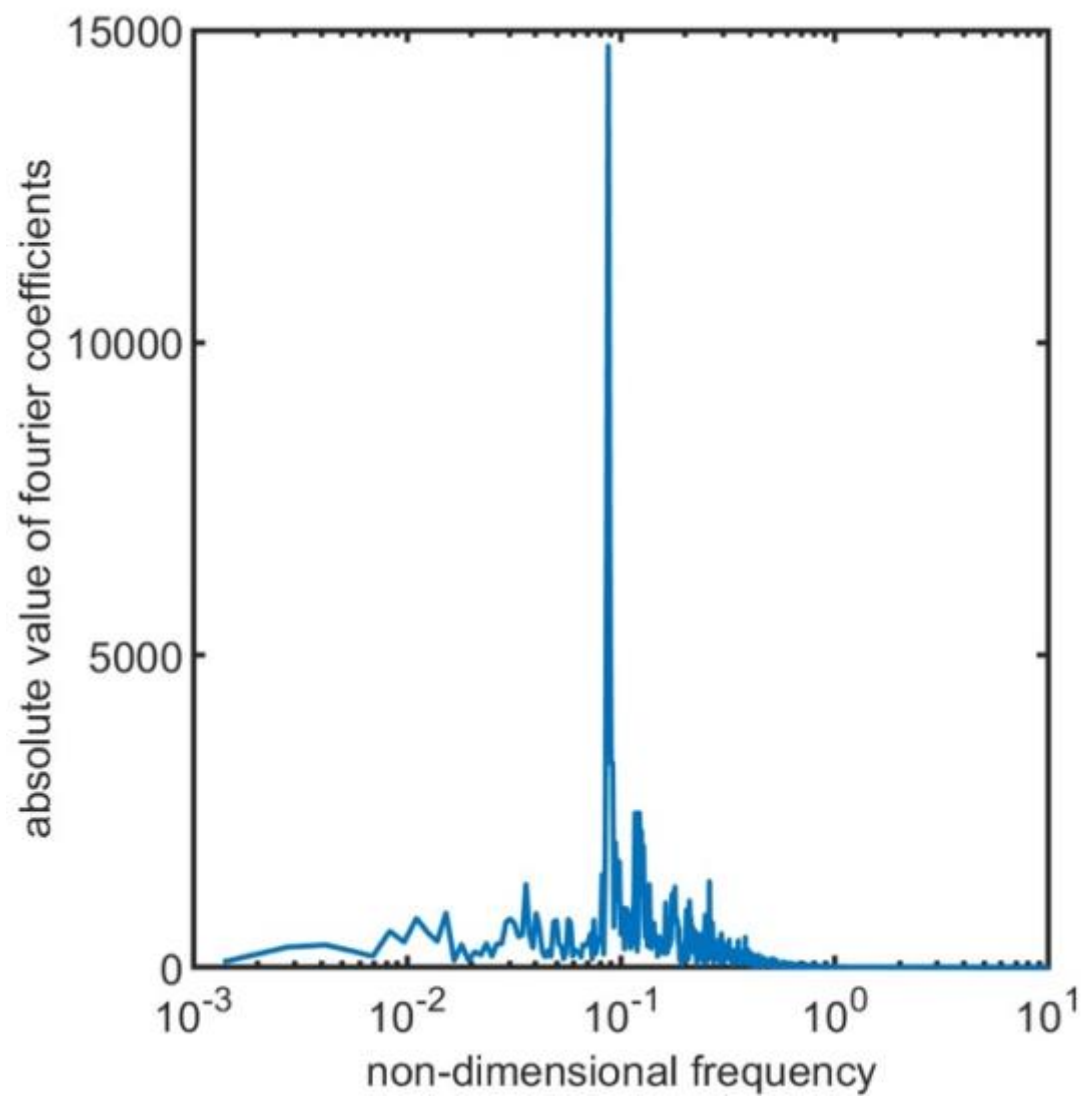

**FIG. S4.** Frequency spectrum of the drag force acting on obstacle (3,3) in figure 5(d) at *Re* = 100 shows the presence of a dominant frequency corresponding to vortex shedding as well as non-zero contributions at other frequencies arising from turbulence. Non-dimensional frequency is calculated as $fd/u_m$, where $f$ is the frequency in Hz.

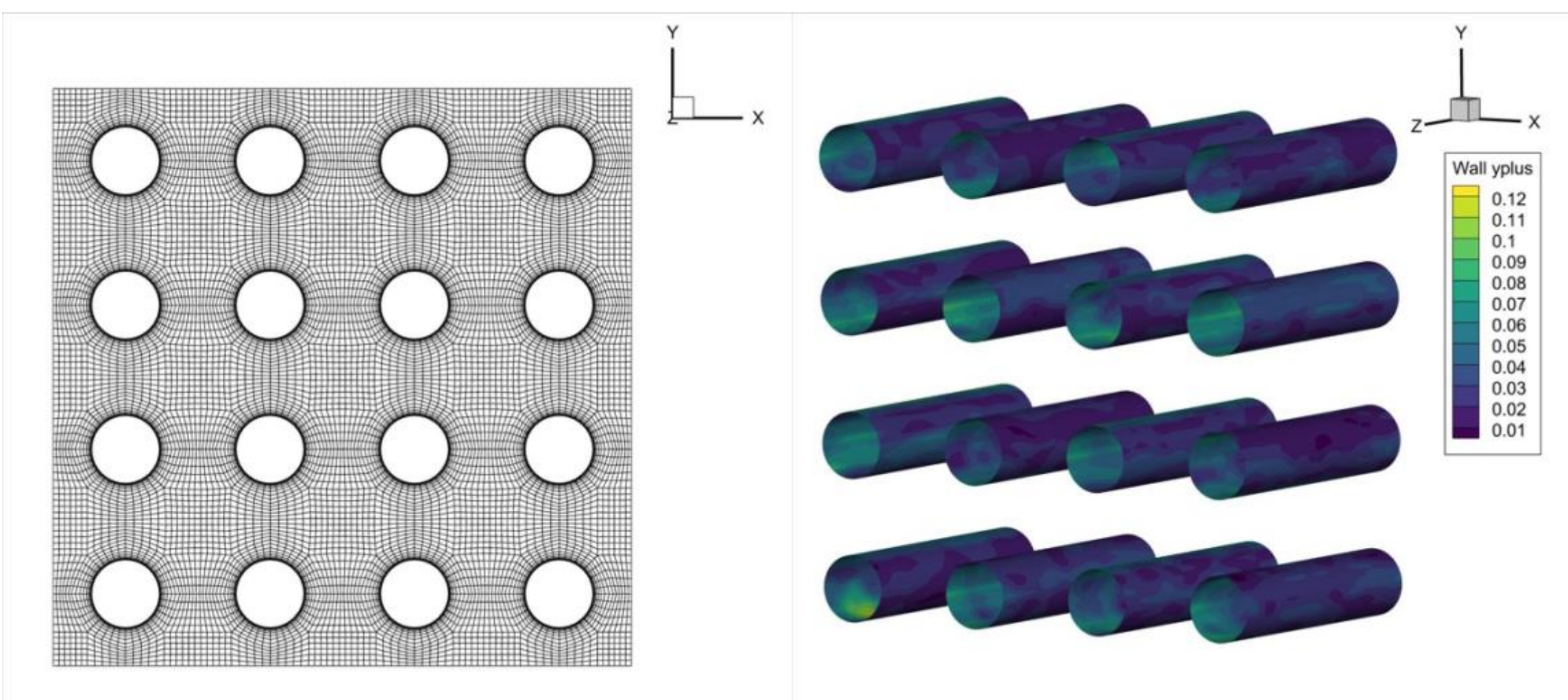

**FIG. S5.** Computational grid used for large eddy simulation with near wall grid refinement that produces $\Delta y^+$ values consistently smaller than 1 indicating that near-wall shear layers have adequate number of grid cells to capture relevant flow physics.